%% file: polar_wom.tex
\documentclass[peerreview, 11pt]{IEEEtran}
\usepackage{graphics}
\usepackage{epsfig}

\usepackage{subfigure}
\usepackage{graphics}

\usepackage{amsmath}
\usepackage{amsfonts}
\usepackage{mathtools}
\usepackage{dsfont}
\usepackage{amssymb}
\usepackage{color}
\definecolor{Blue}{rgb}{0.3,0.3,0.9}
\usepackage{stfloats}

\input{shortcuts.tex}

\begin{document}

\title{Polar Write Once Memory Codes}

\author{David~Burshtein,~\IEEEmembership{Senior Member,~IEEE} and Alona~Strugatski
\thanks{This research was supported by the Israel Science Foundation, grant no. 772/09. This paper was presented in part in ISIT 2012, Boston, MA, July 2012.}
\thanks{D.\ Burshtein is with the school of Electrical Engineering, Tel-Aviv University, Tel-Aviv 69978, Israel (email: burstyn@eng.tau.ac.il).}
\thanks{A.\ Strugatski is with the school of Electrical Engineering, Tel-Aviv University, Tel-Aviv 69978, Israel (email: alonast@gmail.com).}
}

\markboth{Submitted to IEEE Transactions on Information Theory}{Burshtein and Strugatski: Polar write once memory codes}

\maketitle \setcounter{page}{1}

\begin{abstract}
A coding scheme for write once memory (WOM) using polar codes is presented. It is shown that the scheme achieves the capacity region of noiseless WOMs when an arbitrary number of multiple writes is permitted. The encoding and decoding complexities scale as O(N log N) where N is the blocklength. For N sufficiently large, the error probability decreases sub-exponentially in N. The results can be generalized from binary to generalized WOMs, described by an arbitrary directed acyclic graph, using nonbinary polar codes. In the derivation we also obtain results on the typical distortion of polar codes for lossy source coding. Some simulation results with finite length codes are presented.
\end{abstract}


\section{Introduction} \label{sec:introduction}
The model of a write once memory (WOM) was proposed by Rivest and Shamir in~\cite{wom_rivest_shamir}. In write once memories writing may be irreversible in the sense that once a memory cell is in some state it cannot easily convert to a preceding state. Flash memory is an important example since the charge level of each memory cell can only increase, and it is not possible to erase a single memory cell. It is possible to erase together a complete block of cells which comprises a large number of cells, but this is a costly operation and it reduces the life cycle of the device.

Consider a binary write-once memory (WOM) which is comprised of $N$ memory cells. Suppose that we write on the device $t$ times, and denote the number of possible messages in the $i$th write by $M_i$ ($1 \le i \le t$). The number of bits that are written in the $i$th write is $k_i = \log_2 M_i$ and the corresponding code rate is $R_i = k_i / N$.
Let $\bs_l$ denote the $N$ dimensional state vector of the WOM at time (generation) $l$ for $0 \le l \le t$, and suppose that $\bs_0=0$. For $l=1,2,\ldots,t$, the binary message vector is $\ba_l$ ($N R_l$ bits). Given $\ba_l$ and the memory state $\bs_{l-1}$, the encoder computes $\bs_l = \bE_l(\bs_{l-1},\ba_l)$ using an encoding function $\bE_l$ and writes the result $\bs_l$ on the WOM. The WOM constraints can be expressed by $\bs_{l} \ge \bs_{l-1}$ where the vector inequality applies componentwise. Since the WOM is binary, $\bs_{l-1}$ and $\bs_l$ are binary vectors, so that if $s_{l-1,j} = 1$ for some component $j$, then $s_{l,j}=1$. The decoder uses a decoding function $\bD_l$ to compute the decoded message $\hat{\ba}_l = \bD_l(\bs_l)$. The goal is to design a low complexity read-write scheme that satisfies the WOM constraints and achieves $\hat{\ba}_l = \ba_l$ for $l=1,2,\ldots,t$ with high probability for any set of $t$ messages $\ba_l, l=1,2,\ldots,t$.
As is commonly assumed in the literature (see e.g.~\cite{yaakobi2012codes} where it is explained why this assumption does not affect the WOM rate), we also assume that the generation number on each write and read is known.

The capacity region of the WOM is~\cite{heegard1985capacity}
\begin{equation}
\label{eq:C_t}
\begin{split}
C_t &= \left\{ (R_1,\ldots,R_t) \in {\mathbb R}_{+}^t \: | \: R_1 < h(\epsilon_1), \right. \\
    &           R_2 < (1-\epsilon_1)h(\epsilon_2),\ldots,R_{t-1} < \left[ \prod_{j=1}^{t-2} (1-\epsilon_j) \right] h(\epsilon_{t-1}), \\
    &  \left.   R_t < \prod_{j=1}^{t-1} (1-\epsilon_j), \mbox{where } 0 \le \epsilon_1,\epsilon_2,\ldots,\epsilon_{t-1} \le 1/2  \right\}
\end{split}
\end{equation}
(${\mathbb R}_{+}^t$ denotes a $t$-dimensional vector with positive elements; $h(x) = -x \log_2 x - (1-x)\log_2(1-x)$ is the binary entropy function). Note that this is both the zero-error capacity region and the $\epsilon$-error capacity region (see the comment after the statement of Theorem 4 in~\cite{heegard1985capacity}).
We also define the maximum average rate,
$$
\overline{C}_t = \sup_{(R_1,\ldots,R_t) \in C_t} \frac{1}{t} \sum_{j=1}^t R_j
$$
The maximum average rate was shown to be~\cite{heegard1985capacity} $\overline{C}_t = \log_2(t+1) / t$. This means that the total number of bits that can be stored on $N$ WOM cells in $t$ writes is $N \log_2 (t+1)$ which is significantly higher than $N$. The maximum fixed rate was also obtained~\cite{heegard1985capacity}. WOM codes were proposed in the past by various authors, e.g. \cite{wom_rivest_shamir}, \cite{cohen1986linear}, \cite{zemor1991error}, \cite{jiang2008joint}, \cite{wu2010low}, \cite{wu2011position}, \cite{yaakobi2012codes}, \cite{shpilka2011} and references therein.
For the case where there are two writes, $t=2$, the method in~\cite{shpilka2011} can approach capacity in polynomial in the blocklength computational complexity. To the best of our knowledge, this was the first solution with this property.

In this work, which is an expanded version of~\cite{bur_str}, we propose a new family of WOM codes based on polar codes~\cite{arikan2009channel}. The method relies on the fact that polar codes are asymptotically optimal for lossy source coding~\cite{korada2010polar} and can be encoded and decoded efficiently ($O(N\log N)$ operations where $N$ is the blocklength).
We show that our method can achieve any point in the capacity region of noiseless WOMs when an arbitrary number of multiple writes is permitted. The encoding and decoding complexities scale as $O(N\log N)$. For $N$ sufficiently large, the error probability is at most $2^{-N^\beta}$ for any $0 < \beta < 1/2$. We demonstrate that this method can be used to construct actual practical WOM codes. We also show that our results also apply to generalized WOMs, described by an arbitrary directed acyclic graph (DAG), using nonbinary polar codes. In the derivation we also obtain results on the typical distortion of polar codes for lossy source coding.

Recently, another WOM code was proposed~\cite{shpilka2012capacity}, that can approach any point in the capacity region of noiseless WOMs in computational complexity that scales polynomially with the blocklength. On the one hand, the method in~\cite{shpilka2012capacity} is deterministic and guarantees zero error, while our method is probabilistic and only guarantees a vanishing with the blocklength error probability. On the other hand, the method in~\cite{shpilka2012capacity} requires a very long blocklength to closely approach capacity, and it is not clear whether it can be used in practice. In an actual WOM (e.g., flash memory) there is also some channel noise. Hence, there is some small inevitable error.

The rest of this paper is organized as follows. In Section~\ref{sec:polar_background} we provide some background on polar codes for channel and lossy source coding. In Section~\ref{sec:polar_code_extend} we provide extended results on polar codes for lossy source coding that will be required later. In Section~\ref{sec:binary_polar_WOM} we present the new proposed polar WOM code for the binary case and analyze its performance. In Section~\ref{sec:nonbinary_polar_WOM} we present a generalization of our solution to generalized WOMs, described by an arbitrary DAG, using nonbinary polar codes. In Section~\ref{sec:simulations} we present some simulation results. Finally, Section~\ref{sec:conclude} concludes the paper.

\section{Background on Polar codes} \label{sec:polar_background}
In his seminal work~\cite{arikan2009channel}, Arikan has introduced Polar codes for channel coding and showed that they can achieve the symmetric capacity (i.e. the capacity under uniform input distribution) of an arbitrary binary-input channel. In~\cite{sasoglu2009polarization} it was shown that the results can be generalized to arbitrary discrete memoryless channels.
We will follow the notation in~\cite{korada2010polar}.
Let
$G_2 = \left( \begin{array}{cc}
1 & 0 \\
1 & 1 \\
\end{array} \right)$
and let its $n$th Kronecker product be $G_2^{\otimes n}$. Also denote $N=2^n$. Let $\bu$ be an $N$-dimensional binary $\{0,1\}$ message vector, and let $\bx = \bu G_2^{\otimes n}$ where the matrix multiplication is over ${\rm GF}(2)$. Suppose that we transmit $\bx$ over a memoryless binary-input channel with transition probability $W(y \given x)$ and channel output vector $\by$. If $\bu$ is chosen at random with uniform probability then the resulting probability distribution $P(\bu,\bx,\by)$ is given by
\begin{equation}
\label{eq:p_uxy}
P(\bu,\bx,\by) = \frac{1}{2^N} \mathds{1}_{\{\bx = \bu G_2^{\otimes n}\}} \prod_{i=0}^{N-1} W(y_i \given x_i)
\end{equation}
Define the following $N$ sub-channels,
$$
W_N^{(i)}(\by,\bu_0^{i-1} \given u_i)
=
P(\by,\bu_0^{i-1} \given u_i)
=
\frac{1}{2^{N-1}} \sum_{\bu_{i+1}^{N-1}} P(\by \given \bu)
$$
Denote by $I(W)$ the symmetric capacity of the channel $W$ (it is the channel capacity when the channel is memoryless binary-input output symmetric (MBIOS)) and by $Z(W_N^{(i)})$ the Bhattacharyya parameters of the sub-channels $W_N^{(i)}$. In~\cite{arikan2009channel},~\cite{arikan2009rate} it was shown that asymptotically in $N$, a fraction $I(W)$ of the sub-channels satisfy $Z(W_N^{(i)}) < 2^{-2^{n\beta}}$ for any $0 < \beta < 1/2$. Based on this result the following communication scheme was proposed. Let $R$ be the code rate. Denote by $F$ the set of $N(1-R)$ sub-channels with the highest values of $Z(W_N^{(i)})$ (denoted in the sequel as the {\em frozen set}), and by $F^c$ the remaining $N \cdot R$ sub-channels. Fix the input to the sub-channels in $F$ to some arbitrary frozen vector $\bu_F$ (known both to the encoder and to the decoder) and use the channels in $F^c$ to transmit information. The encoder then transmits $\bx = \bu G_2^{\otimes n}$ over the channel. The decoder applies the following successive cancelation (SC) scheme.
For $i=0,1,2,\ldots,N-1$, if $i\in F$ then $\hat{u}_i = u_i$ ($\bu_F$ is common knowledge), otherwise
$$
\hat{u}_i = \left\{
              \begin{array}{ll}
                0 & \hbox{if $L_N^{(i)} > 1$} \\
                1 & \hbox{if $L_N^{(i)} \le 1$}
              \end{array}
            \right.
$$
where
$$
L_N^{(i)}(\by,\bu_0^{i-1}) = \frac{W_N^{(i)}(\by,\hat{\bu}_0^{i-1} \given u_i=0)}{W_N^{(i)}(\by,\hat{\bu}_0^{i-1} \given u_i=1)}
$$
Asymptotically, reliable communication under SC decoding is possible for any $R<I(W)$. The error probability is upper bounded by $2^{-N^\beta}$ for any $\beta<1/2$, and the SC decoder can be implemented in complexity $O(N\log N)$.

Polar codes can also be used for lossy source coding~\cite{korada2010polar}.
Consider a binary symmetric source (BSS), i.e. a random binary vector $\bY$ uniformly distributed over all $N$-dimensional binary vectors. Let $d(\bx,\by)$ be a distance measure between two binary vectors, $\bx$ and $\by$, such that $d(\bx,\by) = \sum_{i=1}^N d(x_i,y_i)$ where $d(0,0)=d(1,1)=0$ and $d(0,1)=d(1,0)=1$. Define a binary symmetric channel (BSC) $W(y \given x)$ with crossover parameter $D$ and construct a polar code with frozen set $F$ that consists of the $(1-R) \cdot N$ sub-channels with the largest values of $Z(W_N^{(i)})$. This code uses some arbitrary frozen vector $\bu_F$ which is known both to the encoder and to the decoder (e.g. $\bu_F=0$) and has rate $R = |F^c|/N$.
\rem{
In fact there is some difference between this code and a polar code constructed for channel coding.
Following the techniques in~\cite{arikan2009rate} it was shown in~\cite{korada2010polar} that asymptotically in $N$, a fraction $1-I(W)$ of the sub-channels of $W$ satisfy $Z(W_N^{(i)}) \ge 1-2^{-2^{n\beta}}$ for any $0 < \beta < 1/2$.
}
Given $\bY=\by$ the SC encoder applies the following scheme. For $i=0,1,\ldots,N-1$, if $i\in F$ then $\hat{u}_i = u_i$, otherwise
\begin{equation}
\label{eq:sc_encoder}
\hat{u}_i = \left\{
              \begin{array}{ll}
                0 & \hbox{w.p. $L_N^{(i)} / (L_N^{(i)} + 1)$} \\
                1 & \hbox{w.p. $1 / (L_N^{(i)} + 1)$}
              \end{array}
            \right.
\end{equation}
(w.p. denotes with probability) The complexity of this scheme is $O(N\log N)$.
Since $\hat{\bu}_F = \bu_F$ is common knowledge, the decoder only needs to obtain $\hat{\bu}_{F^c}$ from the encoder ($|F^c|$ bits). It can then reconstruct the approximating source codeword $\bx$ using $\bx = \hat{\bu} G_2^{\otimes n}$.
Let ${\rm E} d(\bX(\bY),\bY) / N$ be the average distortion  of this polar code (the averaging is over both the source vector, $\bY$, and over the approximating source codeword, $\bX(\bY)$, which is determined at random from $\bY$). Also denote by $R(D)=1-h(D)$ the rate distortion function.
In~\cite{korada2010polar} it was shown, given any $0<D<1/2$, $0 < \delta < 1-R(D)$ and $0<\beta<1/2$, that for $N$ (i.e., $n$) sufficiently large, $R = |F^c|/N = R(D) + \delta$, and any frozen vector $\bu_F$, the polar code with rate $R$ under SC encoding satisfies
\bre
\label{eq:D_N_Avg}
{\rm E} d(\bX(\bY),\bY) / N \le D + O(2^{-N^\beta})
\ere

In fact, as noted in~\cite{korada2010polar}, the proof of~\eqref{eq:D_N_Avg} is not restricted to a BSS and extends to general sources, e.g. a binary erasure source~\cite{korada2010polar}.

\section{Extended results for Polar source codes} \label{sec:polar_code_extend}
Although the result in~\cite{korada2010polar} is concerned only with the average distortion, one may strengthen it by combining it with the strong converse result of the rate distortion theorem in~\cite[p. 127]{csiszar_book}. The strong converse asserts that for any $\delta_1>0$, if $\delta>0$ is chosen sufficiently small and $R < R(D) + \delta$ then $P (d(\bX(\bY),\bY)/N < D-\delta_1)$ can be made arbitrarily small by choosing $N$ sufficiently large. Combining this with~\eqref{eq:D_N_Avg}, we can conclude, for a polar code designed for a BSC($D$), with $R = |F^c|/N \le R(D) + \delta$ and $\delta>0$ sufficiently small, that
\bre
\label{eq:D_N_Prob}
\lim_{N=2^n, n\rightarrow\infty} P (d(\bX(\bY),\bY)/N > D + \delta_2) = 0
\ere
for any $\delta_2>0$.

We now extend the result in~\eqref{eq:D_N_Prob} in order to obtain an improved upper bound estimate (as a function of $N$) on the considered probability. The following discussion is valid for an arbitrary discrete MBIOS, $W(y\given x)$, in~\eqref{eq:p_uxy}. As in~\cite{korada2010polar} we construct a source polar code with frozen set defined by,
\begin{equation}
\label{eq:frozen}
F = \left\{ i\in\{0,...,N-1\} \: : \: Z \left( W_N^{(i)} \right) \geq 1-2\delta_N^2 \right\}
\end{equation}
(note that $F$ depends on $N$, however for simplicity our notation does not show this dependence explicitly) and
\begin{equation}
\label{eq:deltaN}
\delta_N = 2^{-N^\beta} / (2N)
\end{equation}
By~\cite[Theorem 19 and Equation (22)]{korada2010polar} (see also~\cite[Equation (12)]{korada2010polar}),
$$
\lim_{N=2^n,n\rightarrow\infty} |F| / N = 1 - I(W)
$$
Hence, for any $\epsilon>0$, if $N$ is large enough then the rate $R$ of the code satisfies,
$$
R = 1 - |F|/N \le I(W) + \epsilon
$$

Let $\by$ be a source vector produced by a sequence of independent identically distributed (i.i.d.) realizations of $Y$. If $\bu_F$ is chosen at random with uniform probability then the vector $\bu$ produced by the SC encoder (that utilizes~\eqref{eq:sc_encoder}) has a conditional probability distribution given by~\cite{korada2010polar}
\bre
\label{eq:QdefA}
Q(\bu \given \by) = \prod_{i=0}^{N-1} Q(u_i\given \bu_0^{i-1},\by)
\ere
where
\bre
Q(u_i\given \bu_0^{i-1},\by)=
\left\{
  \begin{array}{ll}
    1/2      & \hbox{if $i\in F$} \\
    P(u_i\given \bu_0^{i-1},\by) & \hbox{if $i\in F^c$}
  \end{array}
\right.
\label{eq:Qdef}
\ere
On the other hand, the conditional probability of $\bu$ given $\by$ corresponding to~\eqref{eq:p_uxy} is,
$$
P(\bu\given \by)=\prod_{i=0}^{N-1}P(u_i\given \bu_0^{i-1},\by)
$$

In the sequel we employ standard strong typicality arguments. Similarly to the notation in~\cite[Section 10.6, pp. 325-326]{cover_book},
we define an $\epsilon$-{\em strongly typical} sequence $\bx \in \cX^N$ with respect to a distribution $p(x)$ on the finite set $\cX$, and denote it by $A^{*(N)}_\epsilon(X)$ (or $A^{*(N)}_\epsilon$ for short) as follows. Let $C(a \given \bx)$ denote the number of occurrences of the symbol $a$ in the sequence $\bx$.
Then $\bx \in A^{*(N)}_\epsilon(X)$ if the following two conditions hold. First, for all $a \in \cX$ with $p(a)>0$, ${|C(a \given \bx)/N - p(a)| < \epsilon}$. Second, for all $a \in \cX$ with $p(a)=0$, ${C(a \given \bx)=0}$.
Similarly we define $\epsilon$-strongly typical sequences $\bx,\by \in \cX^N \times \cY^N$ with respect to a distribution $p(x,y)$ on the finite set $\cX \times \cY$, and denote it by $A^{*(N)}_\epsilon(X,Y)$ (or ${A^{*(N)}_\epsilon}$ for short).
We denote by $C(a,b \given \bx,\by)$ the number of occurrences of $a,b$ in $\bx,\by$, and require the following. First, for all $a,b \in \cX \times \cY$ with $p(a,b)>0$,
$|C(a ,b \given \bx,\by)/N - p(a,b)| < \epsilon$. Second, for all $a,b \in \cX \times \cY$ with $p(a,b)=0$, $C(a,b \given \bx,\by)=0$. The definition of $\epsilon$-strong typicality can be extended to more than two sequences in the obvious way.

In our case $\bx=\bx(\bu) \defined \bu G_2^{\otimes n}$. Note that $G_2^{\otimes n}$ is a full rank matrix. Therefore each vector $\bu$ corresponds to exactly one vector $\bx$. We say that $\bu,\by \in A^{*(N)}_\epsilon(U,Y)$ if $\bx(\bu),\by \in A^{*(N)}_\epsilon(X,Y)$ with respect to the probability distribution $p(x,y)=W(y\given x)/2$ (see~\eqref{eq:p_uxy}).
\begin{theorem}
Consider a discrete MBIOS, $W(y\given x)$. Suppose that the input binary random variable $X$ is uniformly distributed (i.e., $X\in\{0,1\}$ w.p. $(1/2,1/2)$), and denote the channel output random variable by $Y$.
Let the source vector random variable $\bY$ be created by a sequence of $N$ i.i.d. realizations of $Y$. Consider a polar code for source coding~\cite{korada2010polar} with block length $N=2^n$ and
a frozen set defined by~\eqref{eq:frozen}-\eqref{eq:deltaN} (whose rate approaches $I(W)$ asymptotically) as described above.
Let $\bU$ be the random variable denoting the output of the SC encoder. Then for any $0 < \beta < 1/2$, $\epsilon>0$ and $N$ (i.e., $n$) sufficiently large,
$\bU,\bY \in A_{\epsilon}^{*(N)}(U,Y)$ w.p. at least $1 - 2^{-N^\beta}$.
\label{th1}
\end{theorem}

Recall that the SC encoder's output $\bu$ has conditional probability distribution $Q(\bu \given \by)$ given by~\eqref{eq:QdefA}-\eqref{eq:Qdef}. Hence, Theorem~\ref{th1} asserts that, for $N$ sufficiently large, $Q\left( A_{\epsilon}^{*(N)}(U,Y) \right) > 1 - 2^{-N^\beta}$.

\beginproof
To prove the theorem we use the following result of~\cite[Lemma 5 and Lemma 7]{korada2010polar},
\begin{equation}
\sum_{\bu,\by} |Q(\bu,\by)-P(\bu,\by)| \leq 2 |F| \delta_N
\label{eq:korada_ineq}
\end{equation}
Hence,
\begin{equation}
\begin{split}
\MoveEqLeft[2] {\left|\sum_{\bu,\by \in {A_\epsilon^{*(N)}}}Q(\bu,\by)-\sum_{\bu,\by \in {A_\epsilon^{*(N)}}}P(\bu,\by)\right|}\\
& \leq {\sum_{\bu,\by \in {A_\epsilon^{*(N)}}}|Q(\bu,\by)-P(\bu,\by) | \leq 2|F|\delta_N}
\label{eq:A}
\end{split}
\end{equation}
In addition we claim the following,
\begin{equation}
\sum_{\bu,\by \in A_\epsilon^{*(N)}}P(\bu,\by) = P\left( A_\epsilon^{*(N)} \right) \geq 1-e^{-N\gamma}
\label{eq:B}
\end{equation}
for some constant $\gamma$ (that can depend on $\epsilon$). We now prove~\eqref{eq:B}.
\begin{align}
\MoveEqLeft[2] {P\left( A_\epsilon^{*(N)} \right) =} \label{eq:EXab}\\
& P\left(\forall a,b \: : \: \left|\frac{1}{N} C(a ,b \given \bX(\bU),\bY)-p(a,b)\right| < \epsilon \right)= \nonumber\\
& 1-P\left(\exists a,b \: : \: \left|\frac{1}{N} C(a ,b \given \bX(\bU),\bY)-p(a,b)\right| \ge \epsilon \right) \nonumber
\end{align}
In the first equality we have used the fact that $p(a,b)=0$ implies $C(a,b \given \bX(\bU),\bY)=0$.
Let $Z$ be a binary $\{0,1\}$ random variable such that $Z_i=1$ if $(X_i(\bU),Y_i)=(a,b)$ and $Z_i=0$ otherwise. Then,
$$
P\left( Z_i = 1 \right) = p(a,b)\quad ,\quad C(a,b \given \bX(\bU),\bY)=\sum_{i=1}^{N} Z_i
$$
Therefore,
\begin{align}
\MoveEqLeft[2] {P\left\{\left|{\frac{1}{N}}C(a,b \given \bX(\bU),\bY)-p(a,b)\right| \ge \epsilon \right\} =} \nonumber\\
& P\left\{\left| \frac{1}{N}\sum_{i=1}^N Z_i - p(a,b) \right| \ge \epsilon \right\} \leq
2e^{-2\epsilon^2N}
\label{eq:C}
\end{align}
where the inequality is due to Hoeffding's inequality (using the fact ${0\leq Z_i\leq 1}$). Hence,
\begin{align*}
\MoveEqLeft[2] {P\left\{\exists a,b \: : \: \left| \frac{1}{N} C(a,b \given \bX(\bU),\bY)-p(a,b)\right| \ge \epsilon \right\} \leq}\\
& 2 |{\cal X||{\cal Y}|}e^{-2 \epsilon^2 N}
\end{align*}
which, together with~\eqref{eq:EXab}, proves \eqref{eq:B}. From \eqref{eq:A}
$$
P\left( A_\epsilon^{*(N)} \right) \leq Q\left( A_\epsilon^{*(N)} \right) + 2|F|\delta_N
$$
Combining this with~\eqref{eq:B} we get
$$
Q\left( A_\epsilon^{*(N)} \right) \geq 1-e^{-N\gamma}-2|F|\delta_N
$$
Recalling the definition of $\delta_N$, \eqref{eq:deltaN}, the theorem follows immediately.
\finproof

Although not needed in the rest of the paper, we can now improve the inequality~\eqref{eq:D_N_Prob} using the following Theorem.
\begin{theorem}
Let $\bY$ be a random vector, uniformly distributed over all $N$-dimensional binary $\{ 0, 1 \}$ vectors. Consider a polar code for source coding~\cite{korada2010polar} designed for a BSC with crossover parameter $D$. Let $\bX(\bY)$ be the reconstructed source codeword given $\bY$. Then for any $\delta>0$, $0 < \beta < 1/2$ and $N$ sufficiently large,
\begin{equation}
\label{eq:MaxD}
Q \left( d(\bX(\bY),\bY)/N \ge D + \delta \right) < 2^{-N^\beta}
\end{equation}
The code rate approaches the rate distortion function, $R(D) = 1-h(D)$, for $N$ sufficiently large.
\label{th2}
\end{theorem}
\beginproof
Since $d(\bX(\bY),\bY) = \sum_{i=0}^N d(X_i,Y_i)$ then,
\begin{equation*}
\begin{split}
\MoveEqLeft[1] { Q \left( d(\bX(\bY),\bY) / N \ge D + \delta \right) =}\\
& Q\left( \left[ C(0,1 \given \bX(\bY),\bY)+C(1,0 \given \bX(\bY),\bY) \right] / N \ge D + \delta \right)
\end{split}
\end{equation*}
Denote by $\cA$, $\cB$ and $\cE$ the events
\begin{align}
\cA &= \left\{ C(0,1 \given \bX(\bY),\bY)/N < D/2 + \delta/2 \right\}\nonumber\\
\cB &= \left\{ C(1,0 \given \bX(\bY),\bY)/N < D / 2 + \delta / 2 \right\}\nonumber\\
\cE &= \left\{ \left[ C(0,1 \given \bX(\bY),\bY) + C(1,0 \given \bX(\bY),\bY) \right] / N < D + \delta \right\}\nonumber
\end{align}
Then for $N$ sufficiently large,
\begin{align}
Q \left( \cE \right) &> Q\left( \cA \cap \cB \right) = 1 - Q\left( \bar \cA \cup \bar \cB \right) \nonumber\\
&\ge 1 - Q\left( \bar \cA \right) - Q\left( \bar \cB \right) > 1-2\cdot2^{-N^\beta}
\label{eq:PrDevent}
\end{align}
The last inequality is due to Theorem~\ref{th1}. This proves~\eqref{eq:MaxD} (since~\eqref{eq:PrDevent} holds for any $0<\beta<1/2$)
\finproof

\section{The proposed polar WOM code} \label{sec:binary_polar_WOM}
Consider the binary WOM problem that was defined in Section~\ref{sec:introduction}.
Given some set of parameters $0 \le \epsilon_1,\epsilon_2,\ldots,\epsilon_{t-1} \le 1/2$, $\epsilon_0 \equiv 0$ and $\epsilon_t \equiv 1/2$, we wish to show that we can construct a reliable polar coding scheme for any set of WOM rates $(R_1,\ldots,R_t) \in {\mathbb R}_{+}^t$ in the capacity region~\eqref{eq:C_t}. That is, the rates satisfy
$$
R_l < \alpha_{l-1}h(\epsilon_l) \quad \forall l=1,2,\ldots,t
$$
where
\bre
\alpha_{l-1} = \prod_{j=0}^{l-1} (1-\epsilon_j)
\label{eq:chan_law_X2}
\ere

\begin{figure}[hbtp]
\begin{center}
\includegraphics[scale=0.75]{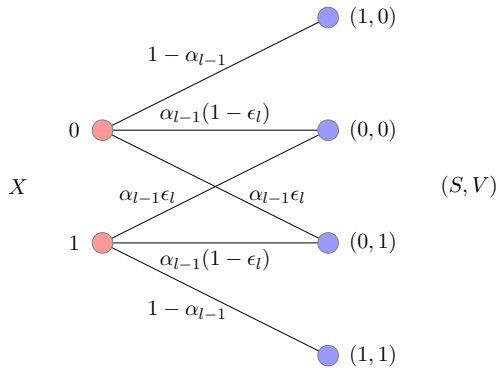}
\caption{The probability transition function of the $l$th channel} \label{fig:binary_channel}
\end{center}
\end{figure}

For that purpose we consider the following $t$ test channels. The input set of each channel is $\{ 0, 1 \}$. The output set is $\{ (0,0), (0,1), (1,0), (1,1) \}$. Denote the input random variable by $X$ and the output by $(S,V)$. The probability transition function of the $l$th channel is defined by,
\bre
P_l \left( (S,V) = (s,v) \given X=x \right) = f_l(s, x \xor v)
\label{eq:chan_law_X0}
\ere
where
\bre
f_l(s,b) =
  \left\{
  \begin{array}{ll}
    \alpha_{l-1}(1-\epsilon_l) & \hbox{if $s=0,b=0$} \\
    \alpha_{l-1}\epsilon_l & \hbox{if $s=0,b=1$} \\
    (1-\alpha_{l-1}) & \hbox{if $s=1,b=0$} \\
    0 & \hbox{if $s=1,b=1$}
  \end{array}
\right.
\label{eq:chan_law_X1}
\ere
This channel is also shown in Figure~\ref{fig:binary_channel}. It is easy to verify that the capacity of this channel is $1-\alpha_{l-1}h(\epsilon_l)$ and that the capacity achieving input distribution is symmetric, i.e., $P(X=0)=P(X=1)=1/2$.

For each channel $l$ we design a polar code with blocklength $N$ and frozen set of sub-channels $F_l$ defined by~\eqref{eq:frozen}. The rate is
\bre
R'_l = 1 - \alpha_{l-1} h(\epsilon_l) + \delta_l
\label{eq:Rl_def}
\ere
where $\delta_l>0$ is arbitrarily small for $N$ sufficiently large.
This code will be used as a source code.

Denote the information sequence by $\ba_1,\ldots,\ba_t$ and the sequence of WOM states by $\bs_0 \equiv 0,\bs_1,\ldots,\bs_t$. Hence $\bs_l = \bE_l(\bs_{l-1},\ba_l)$ and $\hat{\ba}_l=\bD_l(\bs_l)$, where $\bE_l(\bs,\ba)$ and $\bD_l(\bs)$ are the $l$th encoding and decoding functions, respectively, and $\hat{\ba}_1,\ldots,\hat{\ba}_l$ is the retrieved information sequence. We define $\bE_l(\bs,\ba)$ and $\bD_l(\bs)$ as follows.

{\bf Encoding function}, $\hat{\bs} = \bE_l(\bs,\ba)$:
\begin{enumerate}
\item
Let $\bv = \bs \xor \bg$ where $\xor$ denotes bitwise XOR and $\bg$ is a sample from an $N$ dimensional uniformly distributed random binary $\{ 0, 1 \}$ vector. The vector $\bg$ is a common randomness source (dither), known both to the encoder and to the decoder.
\item
Let $y_j = (s_j,v_j)$ and $\by = (y_1,y_2,\ldots,y_N)$. Compress the vector $\by$ using the $l$th polar code with $\bu_{F_l}=\ba_l$. This results in a vector $\bu$ and a vector $\bx = \bu G_2^{\otimes n}$.
\item
Finally $\hat{\bs} = \bx \xor \bg$.
\end{enumerate}

{\bf Decoding function}, $\hat{\ba} = \bD_l(\hat{\bs})$:
\begin{enumerate}
\item
Let $\bx = \hat{\bs} \xor \bg$.
\item
$\hat{\ba} = \left( \bx \left( G_2^{\otimes n} \right)^{-1} \right)_{F_l}$ where $\left( \bz \right)_{F_l}$ denotes the elements of the vector $\bz$ in the set $F_l$.
\end{enumerate}

Note that the information is embedded within the set $F_l$. Hence, when considered as a WOM code, our code has rate $R_l = |F_l| / N = (N-|F^c_l|) / N = 1 - R'_l$, where $R'_l$ is the rate of the polar source code.

For the sake of the proof we slightly modify the coding scheme as follows:
\begin{itemize}
{\setlength\itemindent{7pt} \item[(M1)]
The definition of the $l$th channel is modified such that in~\eqref{eq:chan_law_X1} we use $\epsilon_l-\zeta$ instead of $\epsilon_l$ where $\zeta>0$ will be chosen sufficiently small. We will show that any set of rates $(R_1,\ldots,R_t) \in {\mathbb R}_{+}^t$ that satisfy
$$
R_l < \alpha_{l-1}h(\epsilon_l - \zeta) \quad \forall l=1,2,\ldots,t
$$
is achievable in our scheme. Setting $\zeta$ sufficiently small then shows that any point in the capacity region~\eqref{eq:C_t} is achievable using polar WOM codes.
}
{\setlength\itemindent{7pt} \item[(M2)]
The encoder sets $\bu_{F_l} = \ba_l \xor \bg'_l$ instead of $\bu_{F_l} = \ba_l$, where $\bg'_l$ is $|F_l|$ dimensional uniformly distributed binary (dither) vector known both at the encoder and decoder. In this way, the assumption that $\bu_{F_l}$ is uniformly distributed holds. Similarly, the decoder modifies its operation to $\hat{\ba} = \left( \bx \left( G_2^{\otimes n} \right)^{-1} \right)_{F_l} \xor \bg'_l$.
}
{\setlength\itemindent{7pt} \item[(M3)]
We assume a random permutation of the input vector $\by$ prior to quantization in each polar code. These random permutations are known both at the encoder and decoder.
More precisely, in step 2 the encoder applies the permutation, $\pi$, on $\by$. Then it compresses the permuted $\by$ and obtains some polar codeword.  Finally it applies the inverse permutation, $\pi^{-1}$, on this codeword to produce $\bx$ and proceeds to step 3. The decoder, in the end of step 1, uses the permutation, $\pi$, to permute $\bx$, and then uses this permuted $\bx$ (instead of $\bx$) in step 2.
}
{\setlength\itemindent{7pt} \item[(M4)]
Denote the Hamming weight of the WOM state $\bs_{l}$ after $l$ writes by $\gamma_{l}=w_H(\bs_l)$. Also denote the binomial distribution with $N$ trials and success probability $1-\alpha$ by $B(N,1-\alpha)$, such that $\Upsilon \sim B(N,1-\alpha)$ if for $k=0,1,\ldots,N$,
$
\Pr\left( \Upsilon = k \right) = {N\choose k} (1-\alpha)^k \alpha^{N-k}
$.
After the $l$th write we draw at random a number $\eta$ from the distribution $B(N,1-\alpha_l)$. If $\gamma_l < \eta$ then we flip $\eta-\gamma_l$ elements in $\bs_l$ from $0$ to $1$.
}
\end{itemize}

\begin{theorem}
Consider an arbitrary information sequence $\ba_1,\ldots,\ba_t$ with rates $R_1,R_2,\ldots,R_t$ that are inside the capacity region~\eqref{eq:C_t} of the binary WOM.
For any $0<\beta<1/2$ and $N$ sufficiently large, the coding scheme described above can be used to write this sequence reliably over the WOM w.p. at least $1-2^{-N^\beta}$ in encoding and decoding complexities $O(N\log N)$.
\label{th3}
\end{theorem}

To prove the theorem we need the following lemma\footnote{This Lemma is formulated for the original channel with parameter $\epsilon_l$, and not for the (M1) modified channel with parameter $\epsilon_l-\zeta$.}.
Consider an i.i.d. source $(S,V)$ with the following probability distribution,
\bre
P((S,V)=(s,v)) =
\left\{
  \begin{array}{ll}
    (1-\alpha_{l-1})/2 & \hbox{if $s=1, v=0$} \\
    \alpha_{l-1}/2 & \hbox{if $s=0, v=0$} \\
    \alpha_{l-1}/2 & \hbox{if $s=0, v=1$} \\
    (1-\alpha_{l-1})/2 & \hbox{if $s=1, v=1$}
  \end{array}
\right.
\label{eq:source_distrib}
\ere
Note that this source has the marginal distribution of the output of the $l$th channel defined by~\eqref{eq:chan_law_X0}-\eqref{eq:chan_law_X2} under a symmetric input distribution.
\rem{
Consider the distance measure, $d(x,(s,v)) = \eta(s,x\xor v)$ where $\eta(1,0)=\eta(0,0)=0$, $\eta(0,1)=1$ and $\eta(1,1)=\infty$.
Following~\cite[Section 10.3.1, pp. 307-310]{cover_book}, it can be verified that the rate distortion function, $R(D)$, of this source for $D=\epsilon_l\alpha_{l-1}$ is $R(D)=1-\alpha_{l-1}\epsilon_l$. Hence, when using a polar code with rate $R'_l$ given in~\eqref{eq:Rl_def}, we have
$$
\lim_{N\rightarrow\infty} P \left( d(\bX,(\bS,\bV)) / N > \epsilon_l \alpha_{l-1} + \delta_1 \right) = 0
$$
for any $\delta_1>0$.
}

\begin{lemma}
Consider a polar code designed for the $l$th channel defined by~\eqref{eq:chan_law_X0}-\eqref{eq:chan_law_X2} as described above. The code has rate $R'_l$ defined in~\eqref{eq:Rl_def}, a frozen set of sub-channels, $F_l$, and some frozen vector $\bU_{F_l}$ which is uniformly distributed over all $|F_l|$ dimensional binary vectors. The code is used to encode a random vector $(\bS,\bV)$ drawn by i.i.d. sampling from the distribution~\eqref{eq:source_distrib} using the SC encoder. Denote by $\bX$ the encoded codeword. Then for any $\delta>0$, $0<\beta<1/2$ and $N$ sufficiently large, the following holds w.p. at least $1-2^{-N^\beta}$,
\begin{align}
\label{eq:k_group_size}
\left| \left\{ k \: : \: S_k=0 \mbox{ and } X_k \xor V_k = 1 \right\} \right| <& \left( \alpha_{l-1}\epsilon_l + \delta \right) N \\
\label{eq:empty_k_group}
\left\{ k \: : \: S_k=1 \mbox{ and } X_k \xor V_k = 1 \right\} =& \: \emptyset
\end{align}
\label{lem:1}
\end{lemma}

\beginproof
According to Theorem~\ref{th1}, for $N$ (i.e., $n$) large enough,
$$
(\bX(\bU),(\bS,\bV)) \in A_{\delta/2}^{*(N)}(X,(S,V))
$$
w.p. at least $1-2^{-N^\beta}$. Consider all possible triples $\chi,\xi,\nu$, where $\chi \in \{ 0,1 \}$, $\xi \in \{ 0,1 \}$ and $\nu \in \{ 0,1 \}$.
From the definition of $A_{\delta/2} ^{*(N)}$, if $p(\chi,(\xi,\nu))>0$ then (w.p. at least $1-2^{-N^\beta}$),
\begin{equation}
|C\left( \chi,(\xi,\nu) \given \bX(\bU),(\bS,\bV) \right) / N - p(\chi,(\xi,\nu))| < \delta/2
\label{eq:Strong_typicality_YX}
\end{equation}
and if $p(\chi,(\xi,\nu))=0$ then
\begin{equation}
C\left( \chi,(\xi,\nu) \given \bX(\bU),(\bS,\bV) \right) = 0
\label{eq:Strong_typicality_YX_B}
\end{equation}

In addition, using $P(X=0)= P(X=1)= 1/2$ and the channel definition~\eqref{eq:chan_law_X0}-\eqref{eq:chan_law_X2} we have,
$$
p(0,(0,1)) = p(1,(0,0)) = \alpha_{l-1} \epsilon_l / 2
$$
Combining this with~\eqref{eq:Strong_typicality_YX} we obtain
\begin{align*}
C\left( 0,(0,1) \given \bX(\bU),(\bS,\bV) \right) &< (\alpha_{l-1}\epsilon_l/2+\delta/2)N \\
C\left( 1,(0,0) \given \bX(\bU),(\bS,\bV) \right) &< (\alpha_{l-1}\epsilon_l/2+\delta/2)N
\end{align*}

Hence,
\begin{align*}
\MoveEqLeft[2] {\left| \left\{ k \: : \: S_k=0 \mbox{ and } X_k \xor V_k = 1 \right\} \right| =}\\
& C\left( 0,(0,1) \given \bX(\bU),(\bS,\bV) \right) +\\
& C\left( 1,(0,0) \given \bX(\bU),(\bS,\bV) \right) < (\alpha_{l-1}\epsilon_l+\delta)N
\end{align*}
This proves~\eqref{eq:k_group_size}. Similarly~\eqref{eq:empty_k_group} is due to~\eqref{eq:Strong_typicality_YX_B} since $p(0,(1,1))=p(1,(1,0))=0$ from the definition of the channel.
\finproof

We proceed to the proof of Theorem~\ref{th3}. We denote by $\bS_l,\bS,\hat{\bS},\bV,\bG,\bX$ and $\Gamma_l$ the random variables corresponding to $\bs_l,\bs,\hat{\bs},\bv,\bg,\bx$ and $\gamma_l=w_H(\bs_l)$.

\noindent \textbf{Proof of Theorem~\ref{th3}: }
Note that we only need to prove successful encoding since the WOM is noiseless.

Recall our definition $\Gamma_{l}=w_H(\bS_l)$. Suppose that $\Gamma_{l-1} \sim B(N,1-\alpha_{l-1})$.
Our first claim is that under this assumption, for $\rho>0$ sufficiently small and $N$ sufficiently large, w.p. at least $1-2^{-N^\beta}$, the encoding will be successful and $\Gamma_{l} / N < 1-\alpha_{l}-\rho$.
For notational simplicity we use $\bS$ instead of $\bS_{l-1}$, and $\hat{\bS}$ instead of $\bS_l$.
Considering step 1 of the encoding we see that $\bY = (\bS,\bV)$, after the random permutation described in (M3), can be considered as i.i.d. sampling of the source $(S,V)$ defined in~\eqref{eq:source_distrib} (by the fact that $w_H(\bS)\sim B(N,1-\alpha_{l-1})$, and since $\bG$ is uniformly distributed). Hence, by Lemma~\ref{lem:1} and (M1), the compression of this vector in step 2 satisfies the following for any $\delta>0$ and $N$ sufficiently large w.p. at least $1-2^{-N^\beta}$.
\begin{enumerate}
\item
If $S_k=1$ then $X_k = V_k = S_k \xor G_k = G_k \xor 1$.
\item
For at most $\left[ (\epsilon_l-\zeta)\alpha_{l-1} + \delta \right] N$ components $k$ we have $S_k=0$ and $X_k = V_k \xor 1 = S_k \xor G_k \xor 1 = G_k \xor 1$.
\end{enumerate}
Hence, in step 3 of the encoding, if $S_k=1$ then $\hat{S}_k = X_k \xor G_k = 1$ (i.e. the WOM constraints are satisfied). In addition there are at most $[(\epsilon_l-\zeta)\alpha_{l-1} + \delta]N$ components $k$ for which $S_k=0$ and $\hat{S}_k=1$. Therefore, w.p. at least $1-2^{-N^\beta}$, the vectors $\bS$ and $\hat{\bS}$ satisfy the WOM constraints and
\begin{equation}
\begin{split}
\Gamma_l = w_H(\hat{\bS}) &< [1-\alpha_{l-1}+(\epsilon_l-\zeta)\alpha_{l-1} + 2\delta]N\\
    &= [1-\alpha_l-\zeta\alpha_{l-1} + 2\delta]N
\end{split}
\label{eq:whs}
\end{equation}
(in the first inequality we have used the fact that for $N$ sufficiently large, $\Gamma_{l-1}<(1-\alpha_{l-1}+\delta)N$ w.p. at least $1-e^{-N\epsilon}$ for some $\epsilon>0$ independent of $N$).
Setting $\rho = \zeta\alpha_{l-1}-2\delta$ yields our first claim.

From~\eqref{eq:whs} we know that $\eta$ in (M4) will indeed satisfy the condition $\eta>\Gamma_l$ w.p. at least $1-2^{-N^\beta}$.
The proof of the theorem now follows by using induction on $l$ to conclude that (w.p. at least $1-2^{-N^\beta}$) the $l$th encoding is successful and $\Gamma_l \sim B(N,1-\alpha_{l})$.
The complexity claim is due to the results in~\cite{arikan2009channel}.
\finproof
Notes:
\begin{enumerate}
\item
The test channel in the first write is actually a BSC (since $\alpha_{l-1}=1$ in Figure~\ref{fig:binary_channel}). Similarly, in the last ($t$) write we can merge together the source symbols $(0,0)$ and $(0,1)$ (note that $\epsilon_t=1/2$ so that $X$ and $V$ are statistically independent given $S=0$), thus obtaining a test channel which is a binary erasure channel (BEC).
\item
Consider for example a flash memory device. In practice, the dither, $\bg$, can be determined from the address of the word (e.g. the address is used as a seed value to a random number generator).
\item
In the rare event where an encoding error has occurred, the encoder may re-apply the encoding using another dither vector value. Furthermore, the decoder can realize which value of dither vector should be used in various ways. One possibility is that this information is communicated, similarly to the assumption that the generation number is known. Another possibility is that the decoder will switch to the next value of the dither value upon detecting decoding failure, e.g. by using CRC information. By repeating this procedure of re-encoding upon a failure event at the encoder several times, one can reduce the error probability as much as required.
\end{enumerate}

\section{Generalization to nonbinary polar WOM codes} \label{sec:nonbinary_polar_WOM}
\subsection{Nonbinary polar codes} \label{subsec:nonbinary_polar}
Nonbinary polar codes over a $q$-ary alphabet ($q>2$) for channel coding over arbitrary discrete memoryless channels were proposed in~\cite{sasoglu2009polarization}. Nonbinary polar codes over a $q$-ary alphabet for lossy source coding of a memoryless source were proposed in~\cite{karzand2010qsrc}. First suppose that $q$ is prime. Similarly to the binary case, the codeword $\bx$ of a $q$-ary polar code is related to the $N$-dimensional ($N=2^n$) message vector $\bu$ by the relation $\bx = \bu G_2^{\otimes n}$, where the matrix $G_2^{\otimes n}$ is the same as in the binary case. However, now $\bu \in \cX^N$, $\bx \in \cX^N$ where $\cX = \{ 0,1,\ldots,q-1 \}$. Suppose that we transmit $\bx$ over a memoryless channel with transition probability $W(y \given x)$ and channel output vector $\by$. If $\bu$ is chosen at random with uniform probability over $\cX^N$ then the resulting probability distribution $P(\bu,\bx,\by)$ is given by
\begin{equation}
\label{eq:p_uxy_qary}
P(\bu,\bx,\by) = \frac{1}{q^N} \mathds{1}_{\{\bx = \bu G_2^{\otimes n}\}} \prod_{i=0}^{N-1} W(y_i \given x_i)
\end{equation}
Define the following $N$ sub-channels,
$$
W_N^{(i)}(\by,\bu_0^{i-1} \given u_i)
=
P(\by,\bu_0^{i-1} \given u_i)
=
\frac{1}{q^{N-1}} \sum_{\bu_{i+1}^{N-1}} P(\by \given \bu)
$$
We denote by $I(W)$ and $I(W_N^{(i)})$, respectively, the symmetric capacity parameters of $W$ and $W_N^{(i)}$. In~\cite{sasoglu2009polarization} it was shown that the sub-channels $W_N^{(i)}$ polarize as in the binary case with the same asymptotic polarization rate. The frozen set is chosen similarly to the binary case. Asymptotically, reliable communication under SC decoding is possible for any rate $R<I(W)$. The error probability is upper bounded by $2^{-N^\beta}$ for any $\beta<1/2$, and the decoder can be implemented in complexity $O(N\log N)$.

Nonbinary polar codes were also proposed for lossy source coding~\cite{karzand2010qsrc}. Consider some random variable $Y\in\cY$. For simplicity we assume that $\cY$ is finite. Also denote $\cX=\{0,1,\ldots,q-1\}$. Let the source vector random variable $\bY$ be created by a sequence of $N$ i.i.d. realizations of $Y$. Let $d(x,y)$ be some (finite) distance measure between $x\in\cX$ and $y\in\cY$. Furthermore, for $\bx\in\cX^N$ and $\by\in\cY^N$, we define $d(\bx,\by) = \sum_{i=1}^N d(x_i,y_i)$.
Given some distortion level, $D>0$, let $W(y\given x)$ be the test channel that achieves the symmetric rate distortion bound, $R_s(D)$, (i.e., the rate distortion bound under the constraint that $X$ is uniformly distributed over $\cX$) for the source $Y$ at distortion level $D$. Using that channel, $W(y\given x)$, we construct a polar code with frozen set defined by~\cite{karzand2010qsrc}
\begin{equation}
\label{eq:frozen_qary}
F = \left\{ i\in\{0,...,N-1\} \: : \: I \left( W_N^{(i)} \right) \leq \delta'_N \right\}
\end{equation}
where $\delta'_N = 2^{-N^\beta}$.
Given $\bY=\by$ the SC encoder applies the following scheme. For $i=0,1,\ldots,N-1$, if $i\in F$ then $\hat{u}_i = u_i$, otherwise
$$
\hat{u}_i = m \quad {\rm w.p.} \: \frac{W_N^{(i)}(\by,\hat{\bu}_0^{i-1} \given m)}{\sum_{m'=0}^{q-1} W_N^{(i)}(\by,\hat{\bu}_0^{i-1} \given m')}
$$
The complexity of this scheme is $O(N\log N)$. It was shown~\cite{karzand2010qsrc} that
$$
\lim_{N\rightarrow\infty} |F|/N = 1-I(W)
$$
Hence, for $N$ sufficiently large, the rate of the code, $R=|F^c|/N$, approaches $I(W)=R_s(D)$. Furthermore, for any frozen vector, $\bu_F$,
$$
{\rm E} d(\bX(\bY),\bY) / N \le D + O(2^{-N^\beta})
$$
under SC encoding, where ${\rm E} d(\bX(\bY),\bY) / N$ is the average distortion.

In fact, using the results in~\cite{karzand2010qsrc}, the statements in Section~\ref{sec:polar_code_extend} immediately extend to the nonbinary case. Consider a polar code constructed using some discrete channel $W(y\given x)$ with frozen set defined in~\eqref{eq:frozen_qary}. Suppose that $\bu_F$ is chosen at random with uniform probability. Then, similarly to~\eqref{eq:QdefA}-\eqref{eq:Qdef}, the vector $\bu$ produced by the SC encoder has a conditional probability distribution given by
\bre
\label{eq:QdefA_qary}
Q(\bu \given \by) = \prod_{i=0}^{N-1} Q(u_i\given \bu_0^{i-1},\by)
\ere
where
\bre
Q(u_i\given \bu_0^{i-1},\by)=
\left\{
  \begin{array}{ll}
    1 / q      & \hbox{if $i\in F$} \\
    P(u_i\given \bu_0^{i-1},\by) & \hbox{if $i\in F^c$}
  \end{array}
\right.
\label{eq:Qdef_qary}
\ere
On the other hand, the conditional probability of $\bu$ given $\by$ corresponding to~\eqref{eq:p_uxy_qary} is
$$
P(\bu\given \by)=\prod_{i=0}^{N-1}P(u_i\given \bu_0^{i-1},\by)
$$

Similarly to~\eqref{eq:korada_ineq} above, it was shown in~\cite[Lemma 2 and Lemma 5]{karzand2010qsrc} that
\begin{equation}
\label{eq:korada_ineq_qary}
\sum_{\bu,\by} |Q(\bu,\by)-P(\bu,\by)| \leq |F| \cdot \sqrt{2\log_q e \cdot \delta'_N}
\end{equation}
Combining~\eqref{eq:korada_ineq_qary} with exactly the same arguments that were presented in Theorem~\ref{th1}, yields the following generalization to Theorem~\ref{th1}.
\begin{theorem}
Consider a discrete channel, $W(y\given x)$ where $x\in\cX=\{0,1,\ldots,q-1\}$ and where $q$ is prime. Suppose that the input random variable $X$ is uniformly distributed over $\cX$, and denote the channel output random variable by $Y$. Let the source vector random variable $\bY$ be created by a sequence of $N$ i.i.d. realizations of $Y$. Consider a polar code for source coding~\cite{karzand2010qsrc} with block length $N=2^n$ and a frozen set defined by~\eqref{eq:frozen_qary} (whose rate approaches $I(W)$ asymptotically). Let $\bU$ be the random variable denoting the output of the SC encoder. Then for any $0 < \beta < 1/2$, $\epsilon>0$ and $N$ sufficiently large, $\bU,\bY \in A_{\epsilon}^{*(N)}(U,Y)$ w.p. at least $1 - 2^{-N^\beta}$.
\label{th1_qary}
\end{theorem}

Although not needed in the sequel, Theorem~\ref{th2} also generalizes to the $q$-ary case:
\begin{theorem}
Consider some random variable $Y\in\cY$ and let $\cX = \{ 0,1,\ldots,q-1 \}$ where $q$ is prime. Let the source vector random variable $\bY$ be created by a sequence of $N$ i.i.d. realizations of $Y$. Let $d(x,y)$ be some (finite) distance measure between $x\in\cX$ and $y\in\cY$.
Let $W(y \given x)$ be the test channel that achieves the symmetric rate distortion bound for distance measure $d()$ and some distortion level, $D>0$. Consider a polar code for source coding~\cite{karzand2010qsrc} designed for $W(y\given x)$ as described above. Let $\bX(\bY)$ be the reconstructed source codeword given $\bY$. Then for any $\delta>0$, $0 < \beta < 1/2$ and $N$ sufficiently large,
\begin{equation}
\label{eq:MaxD_qary}
Q \left( d(\bX(\bY),\bY)/N \ge D + \delta \right) < 2^{-N^\beta}
\end{equation}
The code rate approaches the symmetric rate distortion function, $R_s(D)$, for $N$ sufficiently large.
\label{th2_qary}
\end{theorem}

\beginproof
Given some $\delta>0$, we set $\epsilon>0$ sufficiently small and $N$ sufficiently large, thus obtaining
\begin{align*}
Q \left( d(\bX(\bY),\bY)/N \ge D + \delta \right)
&\le Q \left( d(\bX(\bY),\bY)/N \ge D + \delta \: \bigcap \: \: (\bX(\bY), \bY) \in A_\epsilon^{*(N)} \right)\\
&+ Q \left( (\bX(\bY), \bY) \not\in A_\epsilon^{*(N)} \right)\\
&< 2^{-N^\beta}
\end{align*}
where the last inequality is due to Theorem~\ref{th1_qary}, and the fact that if $(\bX(\bY), \bY) \in A_\epsilon^{*(N)}$, for $\epsilon$ sufficiently small and $N$ sufficiently large, then $d(\bX(\bY),\bY)/N < D + \delta$.
\finproof

When $q$ is not prime, the results in this section still apply provided that the polarization transformation is modified as described in~\cite{sasoglu2009polarization}. In each step of the transformation, instead of
$$
x_1 = u_1 + u_2 \quad , \quad x_2 = u_2
$$
we use
$$
x_1 = u_1 + u_2 \quad , \quad x_2 = \pi(u_2)
$$
where $\pi$ is a permutation, chosen at random with uniform probability over $\cX$.

\subsection{The generalized WOM problem} \label{subsec:gen_WOM}
Following~\cite{fu1999capacity}, the generalized WOM is described by a rooted DAG, represented by its set of states (vertices) $\cV$ and by its set of edges $\cE$. The set $\cV = \{ 0,1,\ldots,q-1 \}$ represents the $q$ possible states of each memory cell. We say that there exists a path from state $\theta$ to state $\theta'$ in the WOM, and denote it by $\theta \Rightarrow \theta'$, if, for some $k>0$, there exist vertices $\left\{ \theta=\theta_1,\theta_2,\ldots,\theta_{k-1},\theta_k=\theta' \right\} \in \cV$ such that for $i=1,2,\ldots,k-1$, $\theta_i$ is connected to $\theta_{i+1}$ by an edge in $\cE$ (in particular $\theta \Rightarrow \theta$). The root of the DAG, which represents the initial state of the WOM, is vertex 0. While updating the WOM, only transitions from state $\theta$ to state $\theta'$ where $\theta\Rightarrow \theta'$ are possible. As an example~\cite{fu1999capacity} consider the case where $\cV = \{ 0,1,2,3 \}$ and $\cE = \{ 0 \rightarrow 1, 1 \rightarrow 2, 2 \rightarrow 3 \}$. In this case, we can update a memory cell from state 0 to any other state. We can update from state 1 to either 1, 2, or 3. We can update from state 2 to either 2 or 3. A memory cell in state 3 will remain in this state forever. For two vectors $\btheta, \btheta' \in \cV^N$, we denote by $\btheta \Rightarrow \btheta'$ if and only if $\theta_i \Rightarrow \theta'_i$ for $i=1,2,\ldots,N$. Furthermore, for two random variables $X$ and $Y$ that take values in $\cV$, we denote $X \Rightarrow Y$ if $\Pr \left( X=x, Y=y \right) \ne 0$ only if $x \Rightarrow y$.

The capacity region of the WOM is~\cite{heegard1985capacity}, \cite{fu1999capacity},
\begin{equation}
\label{eq:C_t_gen}
\begin{split}
C_t &= \left\{ (R_1,\ldots,R_t) \in {\mathbb R}_{+}^t \: | \: R_l < H(\Theta_l | \Theta_{l-1}), \right. \\
    &           l=1,2,\ldots,t \mbox{ for some random variables} \\
    &  \left.   0 = \Theta_0 \Rightarrow \Theta_1 \Rightarrow \Theta_2 \ldots \Rightarrow \Theta_t  \right\}
\end{split}
\end{equation}
where $H(\cdot\given\cdot)$ denotes conditional entropy.

Consider some set of random variables such that
\bre
\label{eq:gen_wom_rvs}
0 \equiv \Theta_0 \Rightarrow \Theta_1 \Rightarrow \Theta_2 \ldots \Rightarrow \Theta_t
\ere
Define
\bre
\label{eq:eps_lss}
\epsilon_{l}(\theta,\theta') \defined \Pr \left( \Theta_l = \theta' \given \Theta_{l-1} = \theta \right)
\ere
and
\bre
\label{eq:alpha_lt}
\alpha_l(\theta) \defined \Pr \left( \Theta_l = \theta \right)
\ere
It follows that
\begin{equation}
\alpha_l(\theta') = \sum_{\theta=0}^{q-1} \alpha_{l-1}(\theta) \epsilon_{l}(\theta,\theta')
\label{eq:alpha_ltp}
\end{equation}
and
\bre
\label{eq:HThetGivenThet}
H(\Theta_l \given \Theta_{l-1}) = \sum_{\theta=0}^{q-1} \alpha_{l-1}(\theta) h\left( \left\{ \epsilon_{l}(\theta,\theta') \right\}_{\theta'=0}^{q-1} \right)
\ere
where for a $q$-dimensional probability vector $\bx=(x_0,\ldots,x_{q-1})$ (i.e. $x_i\ge 0$ and $\sum_{i=0}^{q-1} x_i = 1$), the entropy function is defined by
$$
h(\bx) \defined -\sum_{i=0}^{q-1} x_i \log_q x_i
$$
(in this section the base of all the logarithms is $q$ so that code rate is measured with respect to $q$-ary information symbols).

\subsection{The proposed nonbinary polar WOM code} \label{subsec:nonbinary_polar_WOM}
Given a set of random variables $\left\{ \Theta_l \right\}_{l=0}^t$ satisfying~\eqref{eq:gen_wom_rvs}, with parameters defined in~\eqref{eq:eps_lss}-\eqref{eq:alpha_lt}, we wish to show that we can construct a reliable polar WOM coding scheme with WOM rates $(R_1,\ldots,R_t) \in {\mathbb R}_{+}^t$ that satisfy $R_l < H(\Theta_l | \Theta_{l-1})$ for $l=1,2,\ldots,t$, corresponding to the capacity region~\eqref{eq:C_t_gen}.
For that purpose we consider the following $t$ test channels. The input set of each channel is $\{ 0, 1, \ldots, q-1 \}$. The output set is $\{ (s,v) \}_{s,v=0}^{q-1}$. Denote the input random variable by $X$ and the output by $(S,V)$. The probability transition function of the $l$th channel is defined by,
\bre
P_l \left( (S,V) = (s,v) \given X=x \right) = \alpha_{l-1}(s) \epsilon_{l}(s,s+x+v)
\label{eq:chan_law_gen_X0}
\ere
where the additions are modulo $q$.

This channel is symmetric in the following sense~\cite[p. 94]{galbook}. The set of outputs can be partitioned into subsets (the outputs $(s,v)$ with equal value of $s$) such that in the matrix of transition probabilities of each subset, each row (column, respectively) is a permutation of any other row (column). Hence, by~\cite[Theorem 4.5.2]{galbook}, the capacity achieving distribution is the uniform distribution, and the symmetric capacity of this channel is in fact the capacity.
For $X$ uniformly distributed over $\{0,1,\ldots,q-1\}$ we obtain (see Appendix~\ref{AppCapCalc}),
\bre
\label{eq:Ct_qary}
C_l = 1 - H(\Theta_l \given \Theta_{l-1})
\ere

For each channel $l$ we design a polar code with blocklength $N$ and frozen set of sub-channels $F_l$ defined by~\eqref{eq:frozen_qary}. The rate is
\bre
R'_l = 1 - H(\Theta_l \given \Theta_{l-1}) + \delta_l
\label{eq:Rl_gen_def}
\ere
where $\delta_l>0$ is arbitrarily small for $N$ sufficiently large. This code will be used as a source code.

Denote the information sequence by $\ba_1,\ldots,\ba_t$ and the sequence of WOM states by $\bs_0 \equiv 0,\bs_1,\ldots,\bs_t$. Hence $\bs_l = \bE_l(\bs_{l-1},\ba_l)$ and $\hat{\ba}_l=\bD_l(\bs_l)$, where $\bE_l(\bs,\ba)$ and $\bD_l(\bs)$ are the $l$th encoding and decoding functions, respectively, and $\hat{\ba}_1,\ldots,\hat{\ba}_l$ is the retrieved information sequence. We define $\bE_l(\bs,\ba)$ and $\bD_l(\bs)$ as follows. All the additions (and subtractions) are performed modulo $q$.

{\bf Encoding function}, $\hat{\bs} = \bE_l(\bs,\ba)$:
\begin{enumerate}
\item
Let $\bv = \bg - \bs$ where $\bg$ is a sample from an $N$ dimensional uniformly distributed random $q$-ary $\{ 0, 1, \ldots, q-1 \}$ vector. The vector $\bg$ is a common randomness source (dither), known both to the encoder and to the decoder.
\item
Let $y_j = (s_j,v_j)$ and $\by = (y_1,y_2,\ldots,y_N)$. Compress the vector $\by$ using the $l$th polar code with $\bu_{F_l}=\ba_l$. This results in a vector $\bu$ and a vector $\bx = \bu G_2^{\otimes n}$.
\item
Finally $\hat{\bs} = \bx + \bg$.
\end{enumerate}

{\bf Decoding function}, $\hat{\ba} = \bD_l(\hat{\bs})$:
\begin{enumerate}
\item
Let $\bx = \hat{\bs} - \bg$.
\item
$\hat{\ba} = \left( \bx \left( G_2^{\otimes n} \right)^{-1} \right)_{F_l}$.
\end{enumerate}

As in the binary case, the information is embedded within the set $F_l$. Hence, when considered as a WOM code, our code has rate $R_l = |F_l| / N = (N-|F^c_l|) / N = 1 - R'_l$, where $R'_l$ is the rate of the source code.

For the sake of the proof we slightly modify the coding scheme as was done above for the binary case (modifications (M1)-(M4)). More precisely,
\begin{itemize}
{\setlength\itemindent{7pt} \item[(M'1)]
The definition of the $l$th channel is modified such that in~\eqref{eq:chan_law_gen_X0} we use $\epsilon'_{l}(s,s+x+v)$ instead of $\epsilon_{l}(s,s+x+v)$. The parameters $\{ \epsilon'_{l}(\theta,\theta') \}_{\theta,\theta'=0}^{q-1}$, $l=1,2,\ldots,t$, are defined as follows:
\begin{equation}
\epsilon'_l(\theta,\theta') = \left\{
                                \begin{array}{ll}
                                  \epsilon_l(\theta,\theta'), & \hbox{if $\theta \ne 0$} \\
                                  \epsilon_l(\theta,\theta') + \zeta, & \hbox{if $\theta = \theta' = 0$} \\
                                  \epsilon_l(\theta,\theta') - \zeta/(q-1), & \hbox{if $\theta = 0$ and $\theta' \ne 0$.}
                                \end{array}
                              \right.
\label{eq:epsp_def}
\end{equation}
for some $\zeta>0$ which will be chosen arbitrarily small.
In order to obtain a valid set of parameters, $\{ \epsilon'_{l}(\theta,\theta') \}_{\theta,\theta'=0}^{q-1}$, we first argue that we can assume, without loss of generality, that
\begin{equation}
\label{eq:eps_0_theta}
\epsilon_l(0,\theta') > 0, \quad \forall \theta'
\end{equation}
Since vertex $0$ is the root of our DAG then $0 \Rightarrow \theta'$, $\forall \theta'$. If the required condition~\eqref{eq:eps_0_theta} is not satisfied then we can slightly shift the probabilities $\epsilon_l(0,\theta')$ such that~\eqref{eq:eps_0_theta} does hold (all the other transition probabilities, $\epsilon_l(\theta,\theta')$ for $\theta\ne 0$, remain the same). Suppose we can prove the theorem for the shifted parameters. That is, we assume that we can prove the theorem for $R_1,\ldots,R_t$ inside the capacity region~\eqref{eq:C_t_gen} defined with the shifted parameters. Then, by continuity arguments, if we make the difference between the original and shifted parameters sufficiently small, this will also prove the theorem for rates $R_1,\ldots,R_t$ inside the capacity region defined with the original parameters $\left\{ \epsilon_l(\theta,\theta') \right\}$.
}
{\setlength\itemindent{7pt} \item[(M'2)]
The encoder sets $\bu_{F_l} = \ba_l + \bg'_l$ ($\mod q$) instead of $\bu_{F_l} = \ba_l$, where $\bg'_l$ is $|F_l|$ dimensional uniformly distributed $q$-ary (dither) vector known both at the encoder and decoder. In this way, the assumption that $\bu_{F_l}$ is uniformly distributed holds. Similarly, the decoder modifies its operation to $\hat{\ba} = \left( \bx \left( G_2^{\otimes n} \right)^{-1} \right)_{F_l} - \bg'_l$ ($\mod q$).
}
{\setlength\itemindent{7pt} \item[(M'3)]
This modification is identical to modification (M3) above.
}
{\setlength\itemindent{7pt} \item[(M'4)]
Denote by $\gamma_{l,m} = \left| \left\{ j \: : \: s_{l,j} = m \right\} \right|$, where $s_{l,j}$ is the $j$th element of the WOM state after $l$ writes, $\bs_l$. Also denote by $\bgamma_l = \left( \gamma_{l,0},\ldots,\gamma_{l,q-1} \right)$. Let the multinomial distribution with $N$ trials and probabilities $\alpha(0),\ldots,\alpha(q-1)$, $0 \le \alpha(s)\le 1$ and $\sum_{s=0}^{q-1} \alpha(s) = 1$, be denoted by $M(N,\alpha(0),\ldots,\alpha(q-1))$. Then $\bUpsilon \sim M(N,\alpha(0),\ldots,\alpha(q-1))$ if for $k_0,\ldots,k_{q-1} \in [0,N]$ such that $k_0+\ldots+k_{q-1} = N$,
$
\Pr\left( \bUpsilon = (k_0,\ldots,k_{q-1}) \right) = {N\choose k_0,\ldots,k_{q-1}} \prod_{s=0}^{q-1} \alpha(s)^{k_i}
$.
After the $l$th write we draw at random a vector $(\eta_0,\ldots,\eta_{q-1})$ from the distribution $M(N,\alpha_{l}(0),\ldots,\alpha_{l}(q-1))$. If $\gamma_{l,m} < \eta_m$, $\forall m=1,2,\ldots,q-1$, then, $\forall m=1,2,\ldots,q-1$, we flip $\eta_m-\gamma_{l,m}$ elements in $\bs_l$ from $0$ to $m$.
}
\end{itemize}

\begin{theorem}
Consider an arbitrary information sequence $\ba_1,\ldots,\ba_t$ with rates $R_1,R_2,\ldots,R_t$ that are inside the capacity region~\eqref{eq:C_t_gen} of the $q$-ary WOM.
For any $0<\beta<1/2$ and $N$ sufficiently large, the coding scheme described above can be used to write this sequence reliably over the WOM w.p. at least $1-2^{-N^\beta}$ in encoding and decoding complexities $O(N\log N)$.
\label{th3_qary}
\end{theorem}

To prove the theorem we need the following lemma\footnote{This lemma is formulated for the original channel with parameters $\epsilon_{l}(\theta,\theta')$ (and not for the (M'1) modified channel with parameters $\epsilon'_{l}(\theta,\theta')$).}.
Consider an i.i.d. source $(S,V)$ with the following probability distribution,
\bre
P((S,V)=(s,v)) =
\frac{1}{q} \alpha_{l-1}(s)
\;.
\label{eq:source_distrib_gen}
\ere
Note that this source has the marginal distribution of the output of the $l$th channel defined by~\eqref{eq:chan_law_gen_X0} under a symmetric input distribution.

\begin{lemma}
Consider a $q$-ary polar code designed for the $l$th channel defined by~\eqref{eq:chan_law_gen_X0} as described above. The code has rate $R'_l$ defined in~\eqref{eq:Rl_gen_def}, a frozen set of sub-channels, $F_l$, and some frozen vector $\bU_{F_l}$ which is uniformly distributed over all $|F_l|$ dimensional $q$-ary vectors. The code is used to encode a random vector $(\bS,\bV)$ drawn by i.i.d. sampling from the distribution~\eqref{eq:source_distrib_gen} using the SC encoder. Denote by $\bX$ the encoded codeword. Then for any $\delta>0$, $0<\beta<1/2$ and $N$ sufficiently large, the following holds w.p. at least $1-2^{-N^\beta}$ (in the following expressions, additions are modulo $q$),
\bre
\label{eq:k_group_size_gen}
\left| \left\{ k \: : \: S_k=\xi \mbox{ and } X_k + V_k = \nu \right\} \right| < \left[ \alpha_{l-1}(\xi)\epsilon_{l}(\xi,\xi+\nu) + \delta \right] N
\ere
Furthermore, if $\xi \not\Rightarrow \xi+\nu$ then
\bre
\label{eq:empty_k_group_gen}
\left\{ k \: : \: S_k=\xi \mbox{ and } X_k + V_k = \nu \right\} = \emptyset
\ere
\label{lem:2}
\end{lemma}

\beginproof
According to Theorem~\ref{th1_qary}, for $N$ large enough,
$$
(\bX(\bU),(\bS,\bV)) \in A_{\delta/q}^{*(N)}(X,(S,V))
$$
w.p. at least $1-2^{-N^\beta}$. Consider all possible triples $\chi,\xi,\nu$, where $\chi \in \{ 0,1...,q-1 \}$, $\xi \in \{ 0,1...,q-1 \}$ and $\nu \in \{ 0,1...,q-1 \}$.
From the definition of $A_{\delta/q} ^{*(N)}$ we have (w.p. at least $1-2^{-N^\beta}$),
\begin{equation}
|C\left( \chi,(\xi,\nu) \given \bX(\bU),(\bS,\bV) \right) / N - p(\chi,(\xi,\nu))| < \delta/q
\label{eq:Strong_typicality_YX_gen}
\end{equation}
Furthermore, if $p(\chi,(\xi,\nu))=0$ then~\eqref{eq:Strong_typicality_YX_gen} can be strengthened to
\begin{equation}
C\left( \chi,(\xi,\nu) \given \bX(\bU),(\bS,\bV) \right) = 0
\label{eq:Strong_typicality_YX_B_gen}
\end{equation}
In addition, using $P(X=\chi)= 1/q$ and the channel definition~\eqref{eq:chan_law_gen_X0}, we have
$$
p \left( \chi,(\xi,\nu-\chi) \right) = \frac{1}{q} \alpha_{l-1}(\xi) \epsilon_l(\xi,\xi+\nu)
$$
where here, and in the following expressions, $\nu-\chi$ and $\xi+\nu$ are calculated modulo $q$. Combining this with~\eqref{eq:Strong_typicality_YX_gen} we obtain
$$
C\left( \chi,(\xi,\nu-\chi) \given \bX(\bU),(\bS,\bV) \right) < \left( \alpha_{l-1}(\xi) \epsilon_l(\xi,\xi+\nu) / q+\delta/q \right) N
$$

Hence,
\begin{align*}
\MoveEqLeft[2] {\left| \left\{ k \: : \: S_k=\xi \mbox{ and } X_k + V_k =\nu  \right\} \right| =}\\
&\sum_{\chi=0}^{q-1} C\left( \chi,(\xi,\nu-\chi) \given \bX(\bU),(\bS,\bV) \right)\\
&  < \left( \alpha_{l-1}(\xi) \epsilon_l(\xi,\xi+\nu) +\delta \right) N
\end{align*}
($X_k + V_k$ is also calculated modulo $q$). This proves~\eqref{eq:k_group_size_gen}. Similarly,~\eqref{eq:empty_k_group_gen} is due to~\eqref{eq:Strong_typicality_YX_B_gen} since $\xi \not\Rightarrow \xi + \nu$ implies $\epsilon_l(\xi,\xi+\nu) = 0$, and therefore, from the definition of the channel, we have
$p\left( \chi,(\xi,\nu-\chi) \right) = 0$ for all $\chi$.
\finproof

We proceed to the proof of Theorem~\ref{th3_qary}. We denote by $\bS_l,\bS,\hat{\bS},\bV,\bG,\bX$ and $\bGamma_l$ the random variables corresponding to $\bs_l,\bs,\hat{\bs},\bv,\bg,\bx$ and $\bgamma_l$.

\noindent \textbf{Proof of Theorem~\ref{th3_qary}: }
Note that we only need to prove successful encoding since the WOM is noiseless.

Recall our definition $\Gamma_{l,m} = \left| \left\{ j \: : \: S_{l,j} = m \right\} \right|$. Suppose that $\bGamma_{l-1} \sim M(N,\alpha_{l-1}(0),\ldots,\alpha_{l-1}(q-1))$.
Our first claim is that under this assumption, for $\rho>0$ sufficiently small and $N$ sufficiently large, w.p. at least $1-2^{-N^\beta}$, the encoding will be successful and $\Gamma_{l,m} / N < \alpha_l(m) - \rho$ for $m=1,2,\ldots,q-1$.
For notational simplicity we use $\bS$ instead of $\bS_{l-1}$, and $\hat{\bS}$ instead of $\bS_l$.
Considering step 1 of the encoding we see that $\bY = (\bS,\bV)$, after the random permutation described in (M'3), can be considered as i.i.d. sampling of the source $(S,V)$ defined in~\eqref{eq:source_distrib_gen}. Hence, by Lemma~\ref{lem:2} and (M'1), the compression of this vector in step 2 satisfies the following for any $\delta>0$ and $N$ sufficiently large w.p. at least $1-2^{-N^\beta}$.
\begin{enumerate}
\item
Suppose that $\xi \not\Rightarrow \xi+\nu$. Then $\left\{ k \: : \: S_k=\xi \mbox{ and } X_k + V_k = \nu \right\} = \emptyset$. In addition, $V_k = G_k - S_k$ and $\hat{S}_k = X_k + G_k = X_k + V_k + S_k$. Hence we conclude, under the above assumption, that if $S_k=\xi$ then $\hat{S}_k \ne \xi + \nu$.
\item
For at most $\left[ \alpha_{l-1}(\xi)\epsilon'_l(\xi,\xi+\nu) + \delta \right] N$ components $k$ we have $S_k=\xi$ and $X_k + V_k = \nu$, i.e., $\hat{S}_k=\xi+\nu$.
\end{enumerate}
Hence, the WOM constraints are satisfied, and there are at most $\left[ \alpha_{l-1}(\xi) \epsilon'_l(\xi,\xi+\nu) + \delta \right] N$ components $k$ for which $S_k=\xi$ and $\hat{S}_k = \xi + \nu$. Therefore, w.p. at least $1-2^{-N^\beta}$, the vectors $\bS$ and $\hat{\bS}$ satisfy the WOM constraints and
$$
\Gamma_{l,m} < \sum_\xi \left[ \alpha_{l-1}(\xi) \epsilon'_l(\xi,m) + \delta \right] N
$$
Now, recalling~\eqref{eq:epsp_def}, we obtain for $m\ne 0$
\begin{align}
\Gamma_{l,m} &< \sum_{\xi\ne 0} \left[ \alpha_{l-1}(\xi) \left( \epsilon_l(\xi,m) \right) + \delta \right] N \nonumber\\
&+ \left[ \alpha_{l-1}(0) \left( \epsilon_l(0,m) - \frac{\zeta}{q-1} \right) + \delta \right] N \nonumber\\
&= \left[ \alpha_l(m) - \frac{\zeta \alpha_{l-1}(0)}{q-1} + q\delta \right] N
\label{eq:whs_gen}
\end{align}
where the equality is due to~\eqref{eq:alpha_ltp}. Setting $\rho = \zeta \alpha_{l-1}(0) / (q-1) - q \delta$ (note that $\alpha_l(0)>0$ $\forall l$ due to~\eqref{eq:eps_0_theta}) yields our first claim.

From~\eqref{eq:whs_gen} we know that $\eta_1,\ldots,\eta_{q-1}$ in (M'4) will indeed satisfy the condition $\eta_m > \Gamma_{l,m}$ $\forall m = 1,\ldots,q-1$, w.p. at least $1-2^{-N^\beta}$.
The proof of the theorem now follows by using induction on $l$ to conclude that (w.p. at least $1-2^{-N^\beta}$) the $l$th encoding is successful and $\Gamma_l \sim M(N,\alpha_l(0),\ldots,\alpha_l(q-1))$.
\finproof

\section{Simulation Results} \label{sec:simulations}
\begin{figure}[hbtp]
\begin{center}
\includegraphics[scale=0.9]{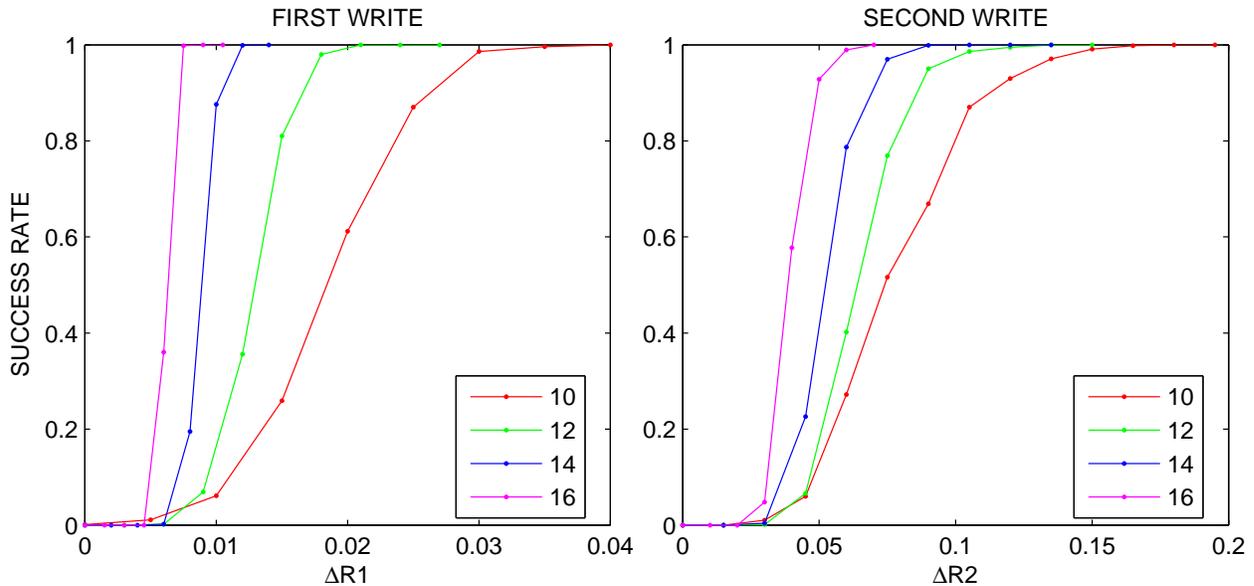}
\caption{The success probabilities of the first (left) and second (right) writes to the binary WOM as a function of the rates loss, $\Delta R_1$ and $\Delta R_2$.}
\label{fig:two_write}
\end{center}
\end{figure}

To demonstrate the performance of our coding scheme for finite length codes we performed experiments with polar WOM codes with $n=10,12,14,16$. Each polar code was constructed using the test channel in Figure~\ref{fig:binary_channel} with the appropriate parameters $\epsilon_l$ and $\alpha_{l-1}$. To learn the frozen set $F_l$ of each code we used the Monte-Carlo approach that was described in~\cite{korada2009polar} (which is a variant of the method proposed by Arikan~\cite{arikan2009channel}).
\rem{
According to this method, we use a random, uniformly distributed, $N$-dimensional information vector, $\bU$. This vector is encoded in order to produce the codeword $\bX = \bU G_2^{\otimes n}$. The codeword $\bX$ is transmitted through the channel in order to produce the channel output, $\bY$, which is the input to the decoder. The decoder is the standard SC decoder, except that for the learning stage the frozen set is empty. The output of the learning procedure is the error probability of each sub-channel, which is used to sort the sub-channels. We preformed 1000 Monte Carlo simulations to learn the sub-channels.
}
Figure~\ref{fig:two_write} describes our experiments with $t=2$ write WOMs designed to maximize the average rate. Using the results in~\cite{heegard1985capacity} we set $\epsilon_1=1/3$. Hence $\alpha_1=2/3$. Each point in each graph was determined by averaging the results of $1000$ Monte-Carlo experiments.
Figure~\ref{fig:two_write} (left) shows the success rate of the first write as a function of the rate loss $\Delta R_1$ compared to the optimum ($R_1=h(1/3)=0.9183$) for each value of $n$. Here success is defined as $w_H(\bs_1)/N \le \epsilon_1$.
Figure~\ref{fig:two_write} (right) shows the success rate of the second write as a function of the rate loss $\Delta R_2$ compared to the optimum ($R_2=2/3$). Here we declare a success if the WOM constraints are satisfied. Each experiment in the second write was performed by using a first write with rate loss of $\Delta R_1=0.01$. For $n=10,12,14$, $\Delta R_1$ should be higher, but this is compensated by using higher values of $\Delta R_2$. As an alternative we could have used a higher rate loss $\Delta R_1$ for $n=10,12,14$, in which case $\Delta R_2$ decreases. In terms of total rate loss both options yielded similar results. We see that for $n=16$ the total rate loss required for successful (with very high probability) first and second write is about $0.08$. As was noted earlier, the success rate can be increased if, upon detecting an encoding error, the encoder repeats the encoding with another dither value.

We have also experimented with a $t=3$ write WOM. We used polar codes with $n=12,14,16$ and set $\epsilon_1=1/4$, $\epsilon_2=1/3$ and $\epsilon_3=1/2$ ($\alpha_1=3/4$ and $\alpha_2=1/2$) to maximize the average rate in accordance with~\cite{heegard1985capacity}. To find the frozen set $F_l$ of each code we used density evolution as described in~\cite{mori2009per} with quantization step $q=0.25$. The maximum average rate is obtained for $R_1=.8113$, $R_2=.6887$ and $R_3=1/2$. The actual information rates are presented in Table~\ref{table:polar_wom_t3}, where in $M$ read/write experiments all information triples were encoded (and decoded) successfully.
\begin{table}
\begin{center}
\begin{tabular}{|c|c|c|c|c|c|c|c|}
\hline
$n$ & $M$ & $\Delta R_1$  & $\Delta R_2$ & $\Delta R_3$ & $R_1$ & $R_2$ & $R_3$ \\
\hline
12  & 10000  & .035  & .05 & .185 &  .776& .639& .315  \\
\hline
14  & 10000  & .02  & .04 & .175 &  .7913& .6487& .325  \\
\hline
16  & 1000  & .02  & .02 & .16 &  .7913& .6687& .34  \\
\hline
\end{tabular}
\end{center}
\caption{The performance of $t=3$ write polar WOMs with $n=12, 14, 16$.}
\label{table:polar_wom_t3}
\end{table}

Finally, we performed experiments with generalized WOMs for the case where the DAG representing the WOM is the following. The vertices are $\cV = \{ 0,1,2 \}$ and the edges are $\cE = \{ 0 \rightarrow 1, 1 \rightarrow 2 \}$. We consider the case where there are two writes, i.e., $t=2$. By~\cite[Proof of Theorem 3.2]{fu1999capacity} the maximum total number of $3$-ary information symbols, $R_1+R_2$, that can be stored in one storage cell of the WOM is $1.6309$. Furthermore, by~\cite[Proof of Theorem 3.2]{fu1999capacity}, the maximum value of $R_1+R_2$ is achieved for $R_1=0.9206$ and $R_2=0.7103$, and the following parameters, $\left\{ \alpha_l(\theta) \right\}_{\theta=0}^2$ and $\left\{ \epsilon_l(\theta,\theta') \right\}_{\theta,\theta'=0}^2$ for $l=0,1$, need to be used in~\eqref{eq:C_t_gen}.
\begin{align*}
& \alpha_0(0)     = 1,   \: \alpha_0(1)=0,         \: \alpha_0(2)=0\\
& \epsilon_0(0,0) = 1/2, \: \epsilon_0(0,1) = 1/3, \: \epsilon_0(0,2) = 1/6\\
& \epsilon_0(1,0) = 0,   \: \epsilon_0(1,1) = 2/3, \: \epsilon_0(1,2) = 1/3\\
& \epsilon_0(2,0) = 0,   \: \epsilon_0(2,1) = 0,   \: \epsilon_0(2,2) = 1\\
& \alpha_1(0)     = 1/2, \: \alpha_1(1)=1/3,       \: \alpha_1(2)=1/6\\
& \epsilon_1(0,0) = 1/3, \: \epsilon_1(0,1) = 1/3, \: \epsilon_1(0,2) = 1/3\\
& \epsilon_1(1,0) = 0,   \: \epsilon_1(1,1) = 1/2, \: \epsilon_1(1,2) = 1/2\\
& \epsilon_1(2,0) = 0,   \: \epsilon_1(2,1) = 0,   \: \epsilon_1(2,2) = 1
\end{align*}

Our scheme uses polar codes with $q=3$ and $n=10,12,14$. Using the above parameters, $\left\{ \alpha_0(\theta) \right\}_{\theta=0}^2$ and $\left\{ \epsilon_0(\theta,\theta') \right\}_{\theta,\theta'=0}^2$, in~\eqref{eq:chan_law_gen_X0}, we obtain the following definition of the first test channel
\begin{align*}
P_1((S,V) = (s,v) \given X=0) &= \left\{
                   \begin{array}{ll}
                     1/2 & \hbox{if $(s,v)=(0,0)$} \\
                     1/3 & \hbox{if $(s,v)=(0,1)$} \\
                     1/6 & \hbox{if $(s,v)=(0,2)$}
                   \end{array}
                 \right.
\\
P_1((S,V) = (s,v) \given X=1) &= \left\{
                   \begin{array}{ll}
                     1/3 & \hbox{if $(s,v)=(0,0)$} \\
                     1/6 & \hbox{if $(s,v)=(0,1)$} \\
                     1/2 & \hbox{if $(s,v)=(0,2)$}
                   \end{array}
                 \right.
\\
P_1((S,V) = (s,v) \given X=2) &= \left\{
                   \begin{array}{ll}
                     1/6 & \hbox{if $(s,v)=(0,0)$} \\
                     1/2 & \hbox{if $(s,v)=(0,1)$} \\
                     1/3 & \hbox{if $(s,v)=(0,2)$}
                   \end{array}
                 \right.
\end{align*}
Similarly, the second test channel is given by
$$
P_2((S,V) = (s,v) \given X=x) =
\left\{
  \begin{array}{ll}
    1/6 & \hbox{if $s=0$} \\
    1/6 & \hbox{if $s=1$ and $x+v \mod 3 \ne 2$} \\
    1/6 & \hbox{if $s=2$ and $x+v \mod 3 = 0$} \\
    0   & \hbox{otherwise.}
  \end{array}
\right.
$$

We see that given $S=0$, $V$ and $X$ are statistically independent. Hence we can simplify this channel by merging the three output symbols, $(0,0)$, $(0,1)$ and $(0,2)$, into one symbol.
To learn the frozen set $F_l$ of each code we used the Monte-Carlo approach that was described in~\cite{korada2009polar}.

\begin{figure}[hbtp]
\begin{center}
\includegraphics[scale=1.2]{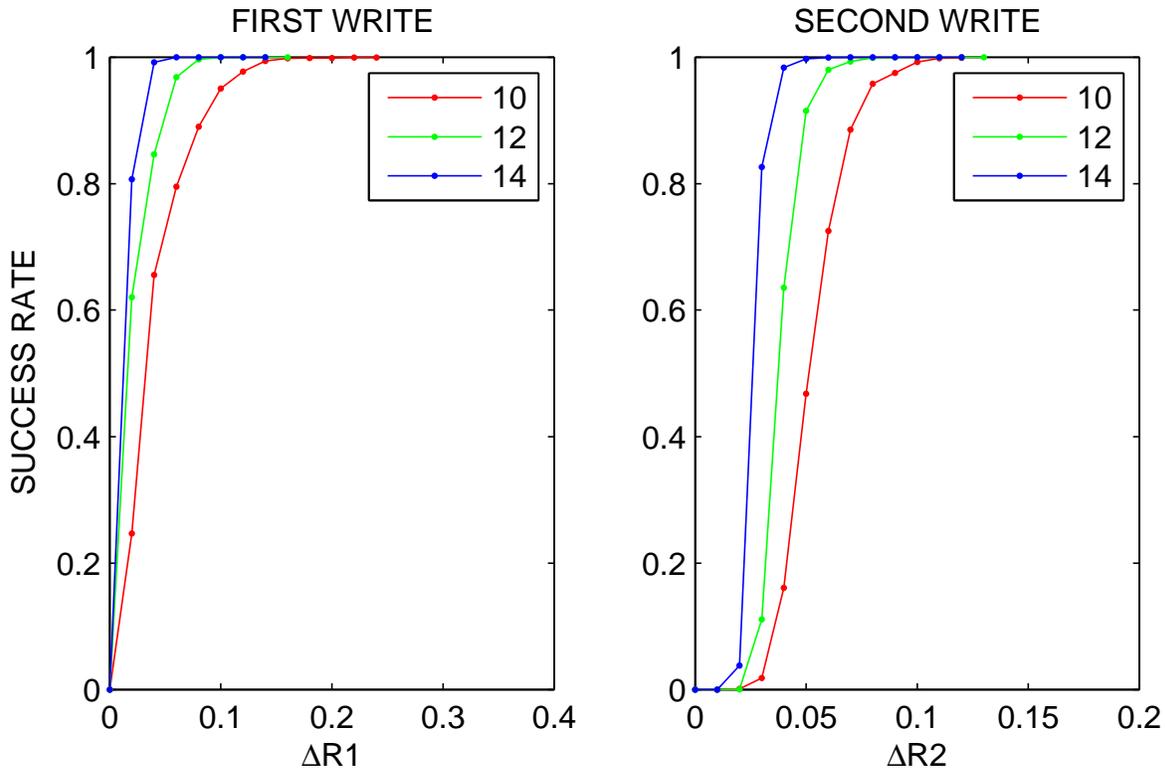}
\caption{The success probabilities of the first (left) and second (right) writes to the $q$-ary WOM for $q=3$ as a function of the rates loss, $\Delta R_1$ and $\Delta R_2$.} \label{fig:q_fig}
\end{center}
\end{figure}

Figure~\ref{fig:q_fig} presents the success rate of the first and second writes as a function of the rates loss $\Delta R_1$ and $\Delta R_2$ compared to the optimal rates, $R_1=0.9206$ and $R_2=0.7103$. This is shown for polar codes with $n=10$, $n=12$ and $n=14$. Each point in the graph was obtained by averaging the results of 10,000 Monte-Carlo experiments. In the first write we declare a success if the fraction of '1' ('2' respectively) in $\bs_1$ is less than or equal to $\alpha_1(1)$ ($\alpha_1(2)$, respectively). In the second write we declare a success if all the WOM constraints are satisfied. Each experiment in the second write was preformed by using $R_1=0.9206$ (i.e., $\Delta R_1 = 0$).

\section{Conclusion} \label{sec:conclude}
We have presented a new family of WOM codes based on the recently proposed polar codes. These codes achieve the capacity region of noiseless WOMs when an arbitrary number of multiple writes is permitted. The encoding and decoding complexities scale as $O(N \log N)$ where $N$ is the blocklength. For $N$ sufficiently large the error probability decreases sub-exponentially in $N$. The results apply both for binary and for generalized WOMs, described by an arbitrary DAG.

There are various directions in which our work can be generalized. The first is the design of codes for noisy WOMs. It should be noted that there are various models for noisy WOMs. The capacity region of the most general model, proposed in~\cite{heegard1985capacity}, is yet unknown. However, for certain special cases~\cite{heegard1985capacity}, \cite{wang2011sum}, the maximum average rate, and in some cases even the capacity region are known. The achievable rate region of some noisy WOM models, presented in~\cite{heegard1985capacity}, \cite{wang2011sum}, are based on coding for Gelfand-Pinsker (GP) side information channels. Hence, in this case one may wish to consider the results in~\cite{korada2010polar} for polar coding over a binary side information channel, and combine them with our method.

Another possibility for further research is the consideration of other codes or decoding methods in our scheme. For example, instead of polar source codes, one may consider low-density generating-matrix (LDGM) codes that were shown in the past to be useful for lossy source coding. Even if polar codes are kept in our scheme, it may be possible to improve performance by using iterative encoding combined with decimation instead of SC encoding. This is due to the fact that iterative decoding usually yields better results compared to SC decoding of polar codes~\cite{arikan2009channel}, \cite{korada2009polar}. One may also consider using list decoding of polar codes as proposed in~\cite{tal2011list}.

\appendices
\section{Derivation of~\eqref{eq:Ct_qary} \label{AppCapCalc}}
For $X$ uniformly distributed over $\{0,1,\ldots,q-1\}$ we have $H(X) = \log_q q = 1$. In addition,
$$
H(X\given (S,V))=\sum_{s,v} P\left( (S,V)=(s,v) \right) H(X\given (S,V)=(s,v))
$$
where
$$
H(X\given (S,V)=(s,v)) = -\sum_x P(X=x\given (S,V)=(s,v)) \log\left( P(X=x \given (S,V)=(s,v)) \right)
$$
Now,
\begin{align*}
P(X=x \given (S,V)=(s,v)) &= \frac{P((S,V)=(s,v) \given X=x) P(X=x)}{P((S,V)=(s,v))} \\
&= \frac{\frac{1}{q} \alpha_{l-1}(s) \epsilon_{l}(s,s+v+x)}{\sum_{x'=0}^{q-1}\frac{1}{q}\alpha_{l-1}(s) \epsilon_{l}(s,s+v+x')} \\
&= \epsilon_{l}(s,s+v+x)
\end{align*}
Hence,
$$
H(X\given (S,V)=(s,v))=h\left( \left\{ \epsilon_{l}(s,s') \right\}_{s'=0}^{q-1} \right)
$$
In addition,
$$
P\left( (S,V)=(s,v) \right)=\frac{1}{q}\alpha_{l-1}(s)
$$
Hence,
\begin{align*}
H(X\given (S,V)) &= \sum_{s=0}^{q-1} \sum_{v=0}^{q-1} \frac{1}{q}\alpha_{l-1}(s) H \left( X\given (S,V)=(s,v) \right) \\
&= \sum_{s=0}^{q-1} \alpha_{l-1}(s) h\left( \left\{ \epsilon_{l}(s,s') \right\}_{s'=0}^{q-1} \right) \\
&= H\left( \Theta_l \given \Theta_{l-1} \right)
\end{align*}
where the last equality is due to~\eqref{eq:HThetGivenThet}. Thus we conclude that
$$
C_l = H(X) - H(X\given (S,V)) = 1 - H\left( \Theta_l \given \Theta_{l-1} \right)
$$
and we have obtained~\eqref{eq:Ct_qary}.

\bibliographystyle{IEEEtran}
\bibliography{IEEEabrv,bibliography}

\end{document}

%% file: shortcuts.tex
\newcommand{\rem}[1]{}

\newtheorem{lemma}{Lemma}

\newtheorem{theorem}{Theorem}

\newcommand{\bre}{\begin{equation}}
\newcommand{\ere}{\end{equation}}

\newcommand{\ee}\]
\newcommand{\bra}{\begin{eqnarray}}
\newcommand{\era}{\end{eqnarray}}
\newcommand{\bfg}{\begin{figure}[hbtp]}
\newcommand{\efg}{\end{figure}}
\newcommand{\bver}{\begin{verbatim}}
\newcommand{\ever}{\end{verbatim}}
\newcommand{\bit}{\begin{itemize}}
\newcommand{\eit}{\end{itemize}}
\newcommand{\ben}{\begin{enumerate}}
\newcommand{\een}{\end{enumerate}}

\newcommand{\bgamma}{\boldsymbol\gamma}

\newcommand{\btheta}{\boldsymbol\theta}

\newcommand{\given}{\: | \:}

\newcommand{\bGamma}{\boldsymbol\Gamma}
\newcommand{\bUpsilon}{\boldsymbol\Upsilon}

\newcommand{\bg}{{\bf{g}}}
\newcommand{\bG}{{\bf{G}}}
\newcommand{\bE}{{\bf E}}

\newcommand{\baa}{\begin{eqnarray*}}
\newcommand{\eaa}{\end{eqnarray*}}

\newcommand{\bs}{{\bf s}}
\newcommand{\ba}{{\bf a}}

\newcommand{\bX}{{\bf X}}

\newcommand{\bU}{{\bf U}}

\newcommand{\bY}{{\bf Y}}
\newcommand{\bV}{{\bf V}}

\newcommand{\bS}{{\bf S}}

\newcommand{\xor}{\oplus}
\newcommand{\bu}{{\bf u}}
\newcommand{\bv}{{\bf v}}

\newcommand{\bD}{{\bf D}}

\newcommand{\bx}{{\bf x}}
\newcommand{\by}{{\bf y}}
\newcommand{\bz}{{\bf z}}

\newcommand{\cA}{{\cal A}}

\newcommand{\cE}{{\cal E}}

\newcommand{\cX}{{\cal X}}
\newcommand{\cY}{{\cal Y}}

\newcommand{\cB}{{\cal B}}

\newcommand{\cV}{{\cal V}}

\newcommand{\beginproof}{\noindent \textbf{Proof: }  }
\newcommand{\finproof}{\noindent $\Box$\\}

\def\defined{\: {\stackrel{\scriptscriptstyle \Delta}{=}} \: }

\def\defined{\: {\stackrel{\scriptscriptstyle \Delta}{=}} \: }

\newfont{\boldlarge}{msbm10 scaled 1100}

\newcommand{\comment}[1]{}

\newlength{\tmpbigbar}
